# Effects of surface integrity on the mechanics of ultra-thin films

Mohamed Shaat∗

*Department of Mechanical Engineering, Zagazig University, Zagazig 44511, Egypt*

*Department of Mechanical and Aerospace Engineering, New Mexico State University, Las Cruces, NM 88003, USA*

**Abstract**

A detailed formulation for the effects of the surface integrity, *i.e.* surface topography, surface metallurgy, and surface mechanical properties, on the mechanics of ultra-thin films is proposed in the framework of linear elasticity. In this formulation, the ultra-thin film is modeled as a material bulk covered with two altered layers and two rough surfaces as distinct phases. Two versions of the proposed formulation are developed. In the first version, the governing equations are obtained depending on the general form of the surface topography. In the second version, the governing equations are reformulated utilizing the average parameters of the surface topography. In the proposed formulation, measures are incorporated to account for the effects of the surface topography, *i.e.* waviness and roughness, surface metallurgy, *i.e.* altered layer, and surface excess energy on the mechanics of ultra-thin films. A case study for the static bending of clamped-clamped ultra-thin films is analytically solved. A parametric study on the effects of the surface roughness, waviness, altered layers, and film size on the static bending of ultra-thin films is presented. In this study, new size-dependent behaviors are revealed where the mechanics of ultra-thin films can be significantly altered for the small variations in the surface integrity.

## 1. Introduction

Nowadays, nanomaterials are intensively utilized in designing nano-devices that drastically cover various fields of applications including physical, biological, chemical, and medical applications. Many factors may affect the mechanics of nanomaterials. Being highly sensitive for the small changes in their material structures and/or sizes is a crucial factor that should be considered when designing nanomaterials for a prescribed performance. Various studies have been conducted to demonstrate this fact. For example, the material structure and size effects on the performance of nanosensors [Ilic et al., 2004; Mohr et al., 2014; Shaat, 2015; Shaat and Abdelkefi, 2015], the buckling and elastic characteristics of nanowires [Shaat and Abdelkefi, 2016; Shaat and Abdelkefi, 2017a; Silva et al., 2006], and the nonlinear dynamics of MEMS/NEMS [Shaat and Abdelkefi, 2017b] have been investigated.

∗Corresponding author.
*E-mail address:*
shaat@nmsu.edu; shaatscience@yahoo.com
Tel.:
+15756215929

Not only the material structure and size but also the surface integrity may significantly affect the mechanics of nanomaterials. Surface integrity is the characterization of surfaces of materials. It concerns with describing the surface topography, the surface metallurgy, and the surface mechanical properties [Bellows and Tishler, 1970; Astakhov 2010]. The surface topography (*i.e.* surface texture) is the representation of the outermost layer of the material in its real form. In fact, the real surface of a material is featured with many irregularities including waviness, roughness, faults, cracks, etc.. The properties of these irregularities mainly depend on the material's processing. The surface metallurgy, on the other hand, is the representation of the nature of the altered layer (*i.e.* the layer between the outermost layer and the bulk of the material). The altered layer reflects the impacts of the manufacturing process and temperature on the material's properties [Bellows and Tishler, 1970]. The altered layer of a material experiences different stress levels than the material's bulk and possesses a different material structure. Therefore, materials may subject to residual stresses and hardness and fatigue strength changes depending on the features and properties of the altered layer [Bellows and Tishler, 1970].

The effects of the surface integrity on the mechanical properties of materials were discussed in various studies. For instance, effects of the surface integrity on the fatigue properties of metals [Sharman et al., 2001; Dieter, 1988; Zaltin and Field, 1973; Huang and Ren, 1991; Ramulu et al., 2001], fatigue life of thin membranes [Sinnott et al., 1989], and fracture resistance of cemented carbides [Llanes et al., 2004] were investigated. Peressadko et al., [2005] studied the influences of the surface roughness on the adhesion between elastic bodies. Effects of the surface topography on the contact of flat-metallic surfaces were discussed and modeled by Greenwood and Williamson [1966]. Zhang et al. [2007] and Yang et al. [2006] investigated the impacts of the surface roughness on the wettability of rough superhydrophobic surfaces. The influences of the surface roughness on the conductivity [Fishman and Calecki, 1991] and magnetic properties [Li et al., 1998] of metals were investigated. Weissmuller and Duan [2008] and Daun et al., [2009] studied the effects of the surface roughness on the bending, free vibration, and sensitivities of microcantilevers.

Ultra-thin films made of single crystalline nanomaterials have been harnessed for many applications including MEMS and NEMS [Craighead, 2000; Huang, 2008]. The experiment and the theory demonstrated the fact that the mechanics of ultra-thin films strongly depends on the properties of their surfaces [Cammarata, R.C., 1994; Cammarata and Sieradzki, 1989; Muller and Saul, 2004]. Because surfaces form a different phase, their excess energy influences the mechanics and properties of ultra-thin films [He et al., 2004; Sharma and Ganti, 2004; He and Li, 2006; Zhou and Huang, 2004; Shim et al., 2005; Sun and Zhang, 2003; Zhang and Sun, 2004; Guo and Zhao, 2005; Lu et al., 2006; Huang, 2008; Shaat et al., 2013a, 2013b, 2014; Wang et al., 2016; Raghu et al., 2016; Fang and Zhu, 2017; Shaat and Abdelkefi, 2017b]. Effects of the surface excess energy (*i.e.* surface energy and surface stress) on the mechanics of ultra-thin films have been investigated via atomistic [Zhou and Huang, 2004; Shim et al., 2005; Sun and Zhang, 2003; Zhang and Sun, 2004; Guo and Zhao, 2005] and continuum [Lu et al., 2006; Huang, 2008; Lü et al., 2009; Shaat et al., 2013a, 2013b, 2014] models. These studies demonstrated that the effects of the surface excess energy increase with the decrease in the film thickness.

In this study, a detailed formulation for the effects of the surface integrity on the mechanics of ultra-thin films is presented in the framework of linear elasticity. In the context of this formulation, new measures are introduced to account for the effects of the surface topography, *i.e.* waviness and roughness, surface metallurgy, *i.e.* altered layer, and surface excess energy, *i.e.* surface stress and energy, on the mechanics of ultra-thin films. To this end, the ultra-thin film is represented consisting of a material bulk, two altered layers, and two outermost surfaces. Two different profile functions are introduced for the upper and lower surfaces to model their surface topography. To account for the altered layers and the surface excess energy effects, the total strain energy of the ultra-thin film is formed as the sum of the material bulk strain energy, the strain energy of the two altered layers, and the surface excess energy. Two versions of the proposed formulation are presented. First, the ultra-thin film formulation is generally derived depending on the random form of the upper and lower surface profiles. Then, the average parameters of the surface integrity are introduced to reformulate the derived governing equations in a simplified form. In order to demonstrate the effects the surface integrity on the mechanics of ultra-thin films, a selected case study for the static bending a clamped-clamped ultra-thin film is analytically solved. An intensive study on the effects of the surface roughness, surface waviness, altered layers, and film size on the static bending of ultra-thin films is presented.

## 2. Formulation for surface integrity effects on the mechanics of ultra-thin films

Utilizing Hamilton's principle, the governing equations of motion of ultra-thin films with surface integrity are derived in the framework of linear elasticity. To this end, an ultra-thin film consisting of a material bulk, two altered layers, and two outer surfaces is considered, as shown in Figure 1. The material bulk is considered with a thickness $h$. A Cartesian coordinate system, $(x, y, z)$, is defined at the mid-plane of the material bulk, as shown in Figure 1. The two interface surfaces between the material bulk and the two altered layers are located at $z = \pm h/2$. In order to account for the surface topography, *i.e.* waviness and roughness, the profiles of the outermost upper surface $S^+$ and the outermost lower surface $S^-$ of the film are defined as follows:

$$P^+(x,y) = H^+ + \varpi^+(x,y) + \mathcal{R}^+(x,y)$$
$$P^-(x,y) = H^- + \varpi^-(x,y) + \mathcal{R}^-(x,y)$$

(1)

where $P(x,y)$ is the profile of the outermost surface layer. $\varpi(x,y)$ and $\mathcal{R}(x,y)$ denote the profiles the waviness and roughness, respectively. $H$ denotes the average thickness of the altered layer. The superscripts $+$ and $-$ refer to the upper and lower surfaces, $S^+$ and $S^-$, respectively. It should be noted that $\varpi(x,y)$ defines the heights of the waviness with respect to the nominal average surface of the altered layer, as shown in Figure 1. $\mathcal{R}(x,y)$ represents the roughness heights with respect to the waviness. According to these profiles and the defined coordinate system, the coordinates of points located at the upper and lower outermost surfaces of the film can be defined as $(x, y, z = P^\pm(x,y) \pm h/2)$.

In the present study, five elastic domains, $\Omega_B(x, y, \{z | z \in (-h/2, h/2)\})$, $\Omega_A^+(x, y, \{z | z \in (h/2, h/2 + P^+(x,y))\})$, $\Omega_A^-(x, y, \{z | z \in (-h/2 - P^-(x,y), -h/2)\})$, $S^+(x, y, \{z | z = h/2 + P^+(x,y)\})$, and $S^-(x, y, \{z | z = -h/2 - P^-(x,y)\})$ are, respectively, defined for the material bulk, the upper and lower altered layers, and the upper and lower outermost surface layers of the ultra-thin film. The Latin dummy indices, $i$ or $j$, refer to $x$, $y$, and $z$ while the Greek dummy indices, $\alpha$ or $\beta$, refer to $x$ and $y$.

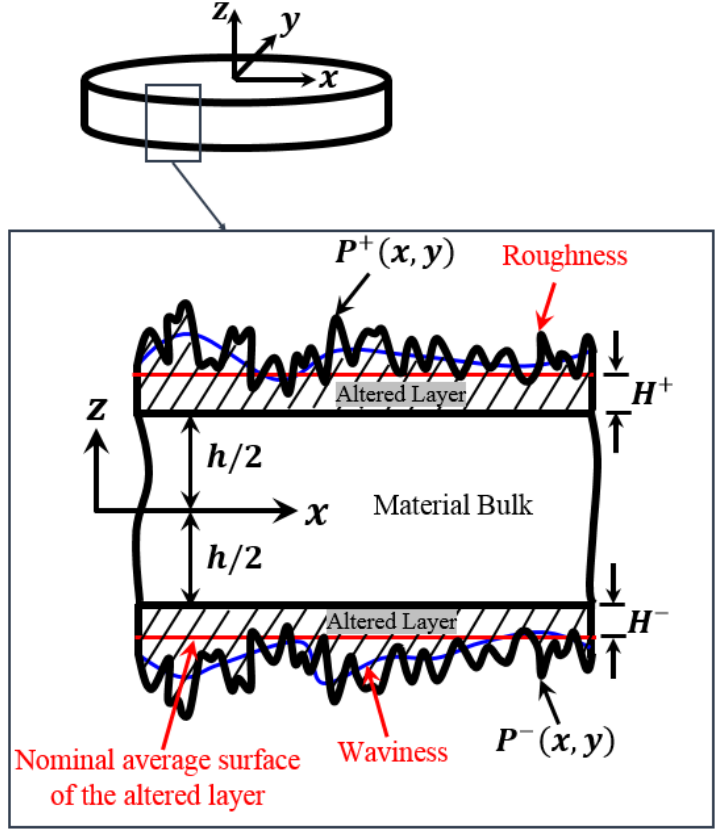

Figure 1: A schematic of an ultra-thin film with surface integrity.

The film is modeled as a Kirchhoff plate; therefore, the displacement field, $u_i$, of a point $(x, y, z)$ belongs to the film can be defined as follows:

$$u_\alpha(x,y,z,t) = -zw_{,\alpha}(x,y,t) \, , \, u_z(x,y,t) = w(x,y,t) \text{ i.e. } \alpha \equiv x,y \tag{2}$$

where $w(x,t)$ is the transverse deflection of the film.

In linear elasticity, the infinitesimal strain tensor, $\varepsilon_{ij} = \frac{1}{2}(u_{i,j} + u_{j,i})$ i.e. $i,j \equiv x,y,z$, is defined as the fundamental measure for deformation. According to the defined displacement field in equation (2), the non-zero strain components that describe the deformation of the ultra-thin film can be obtained as follows:

$$\varepsilon_{\alpha\beta}(x,y,z,t) = -zw_{,\alpha\beta}(x,y,t) \text{ i.e. } \alpha,\beta \equiv x,y \tag{3}$$

For a linear elastic-isotropic film, the constitutive equations of the material bulk can be defined as follows:

$$\sigma^B_{\alpha\beta} = \lambda \varepsilon_{kk}\delta_{\alpha\beta} + 2\mu\varepsilon_{\alpha\beta} \quad \forall z \in \left(-\frac{h}{2}, \frac{h}{2}\right) \tag{4}$$

where $\sigma^B_{\alpha\beta}$ is the stress field within the material bulk domain, $\Omega_B$. $\lambda$ and $\mu$ are the material bulk Lame constants. For thin films, $\lambda = \frac{E\nu}{1-\nu^2}$ and $\mu = \frac{E}{2(1+\nu)}$ where $E$ is the Young's modulus and $\nu$ is the Poisson's ratio.

Similarly, the constitutive equations of the two altered layers can be defined as:

$$\sigma_{\alpha\beta}^{A+} = \lambda_A^+ \varepsilon_{kk} \delta_{\alpha\beta} + 2\mu_A^+ \varepsilon_{\alpha\beta} \; \forall z \in \left(\frac{h}{2}, \frac{h}{2} + P^+(x,y)\right)$$
$$\sigma_{\alpha\beta}^{A-} = \lambda_A^- \varepsilon_{kk} \delta_{\alpha\beta} + 2\mu_A^- \varepsilon_{\alpha\beta} \; \forall z \in \left(-\frac{h}{2} - P^-(x,y), -\frac{h}{2}\right)$$

(5)

where $\sigma_{\alpha\beta}^{A\pm}$ and $\sigma_{\alpha\beta}^{A\pm}$ denote, respectively, the stresses within the upper and lower altered layers of the film. $\lambda_A^\pm$ and $\mu_A^\pm$ are the Lame constants of the upper and lower altered layers, respectively. It should be mentioned that, because of the material processing and the temperature changes, the structure of the altered layer is usually different than the structure of the material bulk [Bellows and Tishler, 1970; Astakhov 2010]. Thus, the altered layer possess different properties and form a distinct phase of the film structure.

Usually, the surfaces of a material form a distinct phase which may affect its mechanics depending on the entire material size [Lu et al., 2006; Huang, 2008; Lü et al., 2009; Shaat et al., 2013a, 2013b, 2014]. These surfaces contribute to the mechanics of the material via an excess surface energy which depends on the surface stress as well as the surface strains. Because ultra-thin films have high surface-to-volume ratios, the excess energy associated with their outermost surfaces should be considered which may significantly affect their mechanics. Therefore, to account for the surface excess energy, constitutive equations are defined for the upper and lower outermost surfaces of the film, $S^+$ and $S^-$, as follows:

$$\sigma_{\alpha\beta}^{S\pm} = \sigma_S^\pm \delta_{\alpha\beta} + \lambda_S^\pm \varepsilon_{kk} \delta_{\alpha\beta} + 2\mu_S^\pm \varepsilon_{\alpha\beta} \; \text{at } z = \pm\frac{h}{2} \pm P^\pm(x,y) \tag{6}$$

where $\sigma_S^\pm$ are, respectively, the surface stresses of the upper and lower surfaces, $S^+$ and $S^-$, where $\lambda_S^\pm$ and $\mu_S^\pm$ denote their surface Lame constants.

To derive the equations of motion of the ultra-thin film, Hamilton's principle is utilized which states that:

$$\int_0^t (\delta T - \delta U + \delta Q) \, dt = 0 \tag{7}$$

where $\delta U$, $\delta T$, and $\delta Q$ are the first variations of the strain energy, the kinetic energy, and the work done, respectively. These quantities can be written in the form:

$$\delta T = -\int_{\Omega_B} (\rho \ddot{u}_i \delta u_i) d\Omega_B - \int_{\Omega_A^+} (\rho_A^+ \ddot{u}_i \delta u_i) d\Omega_A^+ - \int_{\Omega_A^-} (\rho_A^- \ddot{u}_i \delta u_i) d\Omega_A^- \tag{8}$$

$$\delta U = \int_{\Omega_B} (\sigma_{\alpha\beta}^B \delta\varepsilon_{\alpha\beta}) d\Omega_B + \int_{\Omega_A^+} (\sigma_{\alpha\beta}^{A+} \delta\varepsilon_{\alpha\beta}) d\Omega_A^+ + \int_{\Omega_A^-} (\sigma_{\alpha\beta}^{A-} \delta\varepsilon_{\alpha\beta}) d\Omega_A^-$$
$$+ \int_{S^+} (\sigma_{\alpha\beta}^{S+} \delta\varepsilon_{\alpha\beta}) dS^+ + \int_{S^-} (\sigma_{\alpha\beta}^{S-} \delta\varepsilon_{\alpha\beta}) dS^-$$

(9)

$$\delta Q = \int_{\Omega_B} (f_i(x,y) \delta u_i) d\Omega_B + \int_{\Gamma} (t_i \delta u_i) dS \tag{10}$$

where $\rho$ and $\rho_A^{\pm}$ are the mass densities of the material bulk and the upper and lower altered layers, respectively. $f_i(x,y)$ denote the body forces while $t_i$ are the surface tractions.

By substituting equations (8)-(10) into equation (7), the equations of motion of the ultra-thin film can be obtained in the following form (neglecting the rotary inertia terms and the in-plane force components):

$$M_{\alpha\beta,\alpha\beta} + F_z(x,y) - I\ddot{w} = 0 \qquad (11)$$

where

$$F_z(x,y) = \int_{-\frac{h}{2}}^{\frac{h}{2}} f_z(x,y)dz \qquad (12)$$

$$I = \int_{-h/2}^{h/2} \rho\, dz + \int_{h/2}^{h/2+P^+(x,y)} \rho_A^+\, dz + \int_{-h/2-P^-(x,y)}^{-h/2} \rho_A^-\, dz$$

The corresponding boundary conditions at an edge surface, $\Gamma$, can be defined as follows:

$$\begin{aligned} M_{\alpha\beta,\alpha} &= \bar{V}_\alpha \text{ or } w = w^\Gamma \\ M_{\alpha\beta} &= \bar{M}_{\alpha\beta} \text{ or } w_{,\alpha} = w_{,\alpha}^\Gamma \end{aligned} \qquad (13)$$

where $\bar{V}_\alpha$ and $\bar{M}_{\alpha\beta}$ denote the applied shear force and bending moment at edge surface $\Gamma$. $w^\Gamma$ and $w_{,\alpha}^\Gamma$ are prescribed deflection and slop of the film at edge surface $\Gamma$.

For a rectangular film, the boundary conditions (equation (13)) can be explicitly written in the form:

On edge $\Gamma_x \to n_x$ whose normal is $x-$direction ($n_x$):

$$\begin{aligned} M_{xx,x} + M_{xy,y} &= \bar{V}_x \text{ or } w = w^{\Gamma_x} \\ M_{xx} &= \bar{M}_{xx} \text{ or } w_{,x} = w_{,x}^{\Gamma_x} \end{aligned}$$

On edge $\Gamma_y \to n_y$ whose normal is $y-$direction ($n_y$): \qquad (14)

$$\begin{aligned} M_{yy,y} + M_{xy,x} &= \bar{V}_y \text{ or } w = w^{\Gamma_y} \\ M_{yy} &= \bar{M}_{yy} \text{ or } w_{,y} = w_{,y}^{\Gamma_y} \end{aligned}$$

The moment stress resultants introduced in equations (11) and (13) can be defined for the film under consideration as follows:

$$M_{\alpha\beta} = \int_{-h/2}^{h/2} z\sigma_{\alpha\beta}^B\, dz + \int_{h/2}^{h/2+P^+(x,y)} z\sigma_{\alpha\beta}^{A+}\, dz + \int_{-h/2-P^-(x,y)}^{-h/2} z\sigma_{\alpha\beta}^{A-}\, dz + \oint_{z=h/2+P^+(x,y)} z\sigma_{\alpha\beta}^{S+} dz$$
$$+ \oint_{z=-h/2-P^-(x,y)} z\sigma_{\alpha\beta}^{S-} dz \qquad (15)$$

The moment stress resultant can be formed in terms of the deflection by substituting equation (3) into equations (4)-(6) and substituting the result into equation (15), as follows:

$$M_{\alpha\beta} = -\frac{h^3}{12}(\lambda w_{,kk}\delta_{\alpha\beta} + 2\mu w_{,\alpha\beta}) - \frac{1}{3}\left(\left(P^+(x,y) + \frac{h}{2}\right)^3 - \frac{h^3}{8}\right)(\lambda_A^+ w_{,kk}\delta_{\alpha\beta} + 2\mu_A^+ w_{,\alpha\beta})$$

$$-\frac{1}{3}\left(\left(P^-(x,y) + \frac{h}{2}\right)^3 - \frac{h^3}{8}\right)(\lambda_A^- w_{,kk}\delta_{\alpha\beta} + 2\mu_A^- w_{,\alpha\beta})$$

$$-\left(P^+(x,y) + \frac{h}{2}\right)^2 (\lambda_S^+ w_{,kk}\delta_{\alpha\beta} + 2\mu_S^+ w_{,\alpha\beta}) \quad (16)$$

$$-\left(P^-(x,y) + \frac{h}{2}\right)^2 (\lambda_S^- w_{,kk}\delta_{\alpha\beta} + 2\mu_S^- w_{,\alpha\beta}) + (h/2 + P^+(x,y))\sigma_S^+ \delta_{\alpha\beta}$$

$$- (h/2 + P^-(x,y))\sigma_S^- \delta_{\alpha\beta}$$

These moment stress resultants can be explicitly written for rectangular films in the form:

$$M_{xx} = -(D_1(x,y) + D_2(x,y))w_{,xx} - D_2(x,y)w_{,yy} + q(x,y)$$
$$M_{yy} = -(D_1(x,y) + D_2(x,y))w_{,yy} - D_2(x,y)w_{,xx} + q(x,y) \quad (17)$$
$$M_{xy} = -D_1(x,y)w_{,xy}$$

where

$$D_1(x,y) = \mu\left(\frac{h^3}{6}\right) + \frac{2\mu_A^+}{3}\left(\left(P^+(x,y) + \frac{h}{2}\right)^3 - \frac{h^3}{8}\right) + \frac{2\mu_A^-}{3}\left(\left(P^-(x,y) + \frac{h}{2}\right)^3 - \frac{h^3}{8}\right)$$

$$+ 2\mu_S^+\left(P^+(x,y) + \frac{h}{2}\right)^2 + 2\mu_S^-\left(P^-(x,y) + \frac{h}{2}\right)^2$$

$$D_2(x,y) = \lambda\left(\frac{h^3}{12}\right) + \frac{\lambda_A^+}{3}\left(\left(P^+(x,y) + \frac{h}{2}\right)^3 - \frac{h^3}{8}\right) + \frac{\lambda_A^-}{3}\left(\left(P^-(x,y) + \frac{h}{2}\right)^3 - \frac{h^3}{8}\right) \quad (18)$$

$$+ \lambda_S^+\left(P^+(x,y) + \frac{h}{2}\right)^2 + \lambda_S^-\left(P^-(x,y) + \frac{h}{2}\right)^2$$

$$q(x,y) = \sigma_S^+(h/2 + P^+(x,y)) - \sigma_S^-(h/2 + P^-(x,y))$$

By substituting equation (17) into equation (11), the equation of motion can be rewritten in terms of the film deflection as follows:

$$\begin{aligned}
&(D_1(x,y) + D_2(x,y))\left(\frac{\partial^4 w(x,y,t)}{\partial x^4} + 2\frac{\partial^4 w(x,y,t)}{\partial x^2 \partial y^2} + \frac{\partial^4 w(x,y,t)}{\partial y^4}\right) \\
&+ \left[\frac{\partial^2}{\partial x^2}(D_1(x,y) + D_2(x,y)) + \frac{\partial^2}{\partial y^2}D_2(x,y)\right]\frac{\partial^2 w(x,y,t)}{\partial x^2} \\
&+ \left[\frac{\partial^2}{\partial y^2}(D_1(x,y) + D_2(x,y)) + \frac{\partial^2}{\partial x^2}D_2(x,y)\right]\frac{\partial^2 w(x,y,t)}{\partial y^2} \\
&+ 2\left[\frac{\partial^2}{\partial x \partial y}D_1(x,y)\right]\frac{\partial^2 w(x,y,t)}{\partial x \partial y} \\
&+ 2\left[\frac{\partial}{\partial x}(D_1(x,y) + D_2(x,y))\right]\left[\left(\frac{\partial^2}{\partial x^2} + \frac{\partial^2}{\partial y^2}\right)\frac{\partial w(x,y,t)}{\partial x}\right] \\
&+ 2\left[\frac{\partial}{\partial y}(D_1(x,y) + D_2(x,y))\right]\left[\left(\frac{\partial^2}{\partial x^2} + \frac{\partial^2}{\partial y^2}\right)\frac{\partial w(x,y,t)}{\partial y}\right] \\
&- \left[\left(\frac{\partial^2}{\partial x^2} + \frac{\partial^2}{\partial y^2}\right)q(x,y)\right] - F_z(x,y) + I\ddot{w}(x,y,t) = 0
\end{aligned} \qquad (19)$$

where $D_1(x,y)$, $D_2(x,y)$, and $q(x,y)$ are defined in equation (18).

Similarly, by substituting equation (17) into equation (14), the boundary conditions can be written in terms of the film deflection as follows:

On edge $\Gamma_x$:

$$\begin{aligned}
&-(D_1(x,y) + D_2(x,y))(w_{,xxx} + w_{,yyx}) - \left[\frac{\partial}{\partial x}(D_1(x,y) + D_2(x,y))\right]w_{,xx} - \\
&\left[\frac{\partial}{\partial x}D_2(x,y)\right]w_{,yy} - \left[\frac{\partial}{\partial y}D_1(x,y)\right]w_{,xy} + \frac{\partial}{\partial x}q(x,y) = \bar{V}_x \text{ or } w = w^{\Gamma_x} \\
&-(D_1(x,y) + D_2(x,y))w_{,xx} - D_2(x,y)w_{,yy} + q(x,y) = \bar{M}_{xx} \text{ or } w_{,x} = w_{,x}^{\Gamma_x}
\end{aligned}$$

On edge $\Gamma_y$: $\qquad (20)$

$$\begin{aligned}
&-(D_1(x,y) + D_2(x,y))(w_{,yyy} + w_{,xxy}) - \left[\frac{\partial}{\partial y}(D_1(x,y) + D_2(x,y))\right]w_{,yy} - \\
&\left[\frac{\partial}{\partial y}D_2(x,y)\right]w_{,xx} - \left[\frac{\partial}{\partial x}D_1(x,y)\right]w_{,xy} + \frac{\partial}{\partial y}q(x,y) = \bar{V}_y \text{ or } w = w^{\Gamma_y} \\
&-(D_1(x,y) + D_2(x,y))w_{,yy} - D_2(x,y)w_{,xx} + q(x,y) = \bar{M}_{yy} \text{ or } w_{,y} = w_{,y}^{\Gamma_y}
\end{aligned}$$

The derived governing equations (equations (19) and (20)) incorporate measures to account for the surface integrity effects on the various mechanics of ultra-thin films. $P(x,y)$ is introduced to account for the surface topography, *i.e.* waviness and roughness, effects. In addition, $\lambda_A$ and $\mu_A$ are incorporated to account for the effects of the altered layer, which has a distinct material structure. Furthermore, $\sigma_S$, $\lambda_S$, and $\mu_S$ are introduced to measure the effects of the surface energy and surface stress of the outermost layer on the mechanics of the thin film.

**Case 1:** For an ultra-thin film with, both, the upper and the lower surfaces have the same surface integrity parameters, *i.e.* $P^+(x,y) = P^-(x,y) = P(x,y)$, $\lambda_A^+ = \lambda_A^- = \lambda_A$, $\mu_A^+ = \mu_A^- = \mu_A$, $\rho_A^+ = \rho_A^- = \rho_A$, $\sigma_S^+ = \sigma_S^- = \sigma_S$, $\lambda_S^+ = \lambda_S^- = \lambda_S$, and $\mu_S^+ = \mu_S^- = \mu_S$, the equation of motion (19) can be rewritten in the form:

$$\begin{aligned}
&(D_1(x,y) + D_2(x,y))\left(\frac{\partial^4 w(x,y,t)}{\partial x^4} + 2\frac{\partial^4 w(x,y,t)}{\partial x^2 \partial y^2} + \frac{\partial^4 w(x,y,t)}{\partial y^4}\right) \\
&+ \left[\frac{\partial^2}{\partial x^2}(D_1(x,y) + D_2(x,y)) + \frac{\partial^2}{\partial y^2}D_2(x,y)\right]\frac{\partial^2 w(x,y,t)}{\partial x^2} \\
&+ \left[\frac{\partial^2}{\partial y^2}(D_1(x,y) + D_2(x,y)) + \frac{\partial^2}{\partial x^2}D_2(x,y)\right]\frac{\partial^2 w(x,y,t)}{\partial y^2} \\
&+ 2\left[\frac{\partial^2}{\partial x \partial y}D_1(x,y)\right]\frac{\partial^2 w(x,y,t)}{\partial x \partial y} \\
&+ 2\left[\frac{\partial}{\partial x}(D_1(x,y) + D_2(x,y))\right]\left[\left(\frac{\partial^2}{\partial x^2} + \frac{\partial^2}{\partial y^2}\right)\frac{\partial w(x,y,t)}{\partial x}\right] \\
&+ 2\left[\frac{\partial}{\partial y}(D_1(x,y) + D_2(x,y))\right]\left[\left(\frac{\partial^2}{\partial x^2} + \frac{\partial^2}{\partial y^2}\right)\frac{\partial w(x,y,t)}{\partial y}\right] - F_z(x,y) \\
&+ I\ddot{w}(x,y,t) = 0
\end{aligned} \quad (21)$$

with

$$\begin{aligned}
D_1(x,y) &= \mu\left(\frac{h^3}{6}\right) + \frac{4\mu_A}{3}\left(\left(P(x,y) + \frac{h}{2}\right)^3 - \frac{h^3}{8}\right) + 4\mu_S\left(P(x,y) + \frac{h}{2}\right)^2 \\
D_2(x,y) &= \lambda\left(\frac{h^3}{12}\right) + \frac{2\lambda_A}{3}\left(\left(P(x,y) + \frac{h}{2}\right)^3 - \frac{h^3}{8}\right) + 2\lambda_S\left(P(x,y) + \frac{h}{2}\right)^2
\end{aligned} \quad (22)$$

$$I = \int_{-h/2}^{h/2} \rho\, dz + 2\int_{h/2}^{h/2 + P(x,y)} \rho_A\, dz \quad (23)$$

and the corresponding boundary conditions:

On edge $\Gamma_x$:

$$-(D_1(x,y) + D_2(x,y))(w_{,xxx} + w_{,yyx}) - \left[\frac{\partial}{\partial x}(D_1(x,y) + D_2(x,y))\right]w_{,xx} -$$

$$\left[\frac{\partial}{\partial x}D_2(x,y)\right]w_{,yy} - \left[\frac{\partial}{\partial y}D_1(x,y)\right]w_{,xy} = \bar{V}_x \text{ or } w = w^{\Gamma_x}$$

$$-(D_1(x,y) + D_2(x,y))w_{,xx} - D_2(x,y)w_{,yy} = \bar{M}_{xx} \text{ or } w_{,x} = w_{,x}^{\Gamma_x} \quad (24)$$

On edge $\Gamma_y$:

$$-(D_1(x,y) + D_2(x,y))(w_{,yyy} + w_{,xxy}) - \left[\frac{\partial}{\partial y}(D_1(x,y) + D_2(x,y))\right]w_{,yy} -$$

$$\left[\frac{\partial}{\partial y}D_2(x,y)\right]w_{,xx} - \left[\frac{\partial}{\partial x}D_1(x,y)\right]w_{,xy} = \bar{V}_y \text{ or } w = w^{\Gamma_y}$$

$$-(D_1(x,y) + D_2(x,y))w_{,yy} - D_2(x,y)w_{,xx} = \bar{M}_{yy} \text{ or } w_{,y} = w_{,y}^{\Gamma_y}$$

**Case 2:** For an ultra-thin film with super-polished surfaces, *i.e.* $\varpi(x,y) \cong 0$ and $\mathcal{R}(x,y) \cong 0$, the equation of motion (19) can be rewritten in the following form:

$$(D_1 + D_2)\left(\frac{\partial^4 w(x,y,t)}{\partial x^4} + 2\frac{\partial^4 w(x,y,t)}{\partial x^2 \partial y^2} + \frac{\partial^4 w(x,y,t)}{\partial y^4}\right) - F_z(x,y) + I\ddot{w}(x,y,t) \qquad (25)$$
$$= 0$$

With

$$D_1 = \mu\left(\frac{h^3}{6}\right) + \frac{2\mu_A^+}{3}\left(\left(H^+ + \frac{h}{2}\right)^3 - \frac{h^3}{8}\right) + \frac{2\mu_A^-}{3}\left(\left(H^- + \frac{h}{2}\right)^3 - \frac{h^3}{8}\right) + 2\mu_S^+\left(H^+ + \frac{h}{2}\right)^2$$
$$+ 2\mu_S^-\left(H^- + \frac{h}{2}\right)^2$$

$$D_2 = \lambda\left(\frac{h^3}{12}\right) + \frac{\lambda_A^+}{3}\left(\left(H^+ + \frac{h}{2}\right)^3 - \frac{h^3}{8}\right) + \frac{\lambda_A^-}{3}\left(\left(H^- + \frac{h}{2}\right)^3 - \frac{h^3}{8}\right) + \lambda_S^+\left(H^+ + \frac{h}{2}\right)^2 \qquad (26)$$
$$+ \lambda_S^-\left(H^- + \frac{h}{2}\right)^2$$

$$I = \int_{-h/2}^{h/2} \rho \, dz + \int_{h/2}^{h/2+H^+} \rho_A^+ \, dz + \int_{-h/2-H^-}^{-h/2} \rho_A^- \, dz \qquad (27)$$

The boundary conditions (equation (20)) can be reformulated as follows:

On edge $\Gamma_x$:

$$-(D_1 + D_2)(w_{,xxx} + w_{,yyx}) = \bar{V}_x \text{ or } w = w^{\Gamma_x}$$

$$-(D_1 + D_2)w_{,xx} - D_2 w_{,yy} + q = \bar{M}_{xx} \text{ or } w_{,x} = w_{,x}^{\Gamma_x}$$

On edge $\Gamma_y$: $\qquad (28)$

$$-(D_1 + D_2)(w_{,yyy} + w_{,xxy}) = \bar{V}_y \text{ or } w = w^{\Gamma_y}$$

$$-(D_1 + D_2)w_{,yy} - D_2 w_{,xx} + q = \bar{M}_{yy} \text{ or } w_{,y} = w_{,y}^{\Gamma_y}$$

where

$$q = \sigma_S^+(h/2 + H^+) - \sigma_S^-(h/2 + H^-) \qquad (29)$$

It should be mentioned that $H^+$ and $H^-$ are incorporated into the equations of motion (25) and the boundary conditions (28) to account for the surface metallurgy, *i.e.* altered layer, effects. Moreover, the surface stress, $\sigma_S$, may affect the mechanics of ultra-thin films with super-polished surfaces, as shown in equation (28). Thus, the moment boundary condition depends on the film's surface stresses.

**Case 3:** When the surface topography effects are omitted from the derived equations, *i.e.* $P(x,y) = 0$, the equation of motion (19) reduces to the form:

$$(D_1 + D_2)\left(\frac{\partial^4 w(x,y,t)}{\partial x^4} + 2\frac{\partial^4 w(x,y,t)}{\partial x^2 \partial y^2} + \frac{\partial^4 w(x,y,t)}{\partial y^4}\right) - F_z(x,y) + I\ddot{w}(x,y,t) = 0 \qquad (30)$$

with

$$\begin{aligned}
D_1 &= \mu\left(\frac{h^3}{6}\right) + \mu_S^+\left(\frac{h^2}{2}\right) + \mu_S^-\left(\frac{h^2}{2}\right) \\
D_2 &= \lambda\left(\frac{h^3}{12}\right) + \lambda_S^+\left(\frac{h^2}{4}\right) + \lambda_S^-\left(\frac{h^2}{4}\right) \\
q &= \sigma_S^+\left(\frac{h}{2}\right) - \sigma_S^-\left(\frac{h}{2}\right)
\end{aligned} \qquad (31)$$

and the corresponding boundary conditions:

On edge $\Gamma_x$:

$$-(D_1 + D_2)(w_{,xxx} + w_{,yyx}) = \bar{V}_x \text{ or } w = w^{\Gamma_x}$$

$$-(D_1 + D_2)w_{,xx} - D_2 w_{,yy} + q = \bar{M}_{xx} \text{ or } w_{,x} = w_{,x}^{\Gamma_x}$$

(32)

On edge $\Gamma_y$:

$$-(D_1 + D_2)(w_{,yyy} + w_{,xxy}) = \bar{V}_y \text{ or } w = w^{\Gamma_y}$$

$$-(D_1 + D_2)w_{,yy} - D_2 w_{,xx} + q = \bar{M}_{yy} \text{ or } w_{,y} = w_{,y}^{\Gamma_y}$$

Inspecting equations (30)-(32), when the surface topography effects are eliminated, the derived formulation reduces to the formulation of ultra-thin films with surface excess energy effects. This reduced formulation (equations (30)-(32)) matches the formulations of thin Kirchhoff plates with surface effects presented in various studies [Lu et al., 2006; Huang, 2008; Shaat et al., 2014].

**Case 4:** When the surface integrity effects are omitted from the derived equations, i.e. $P(x,y) = 0$, $\sigma_s = 0$, $\lambda_s = 0$, and $\mu_s = 0$, the formulation of the ultra-thin film recovers the classical model of Kirchhoff plate theory.

Next, the derived equations of motion and boundary conditions are reformulated on the bases of the average values of the surface integrity parameters.

## 3. Formulation based on average parameters of surface integrity

It is challenging to derive an analytical solution for the obtained equation of motion (19) in its current form. However, a simplified version of equation (19) can be proposed by introducing $\langle P \rangle$ as the average of the surface profile and $\langle \partial P/\partial x \rangle$ and $\langle \partial P/\partial y \rangle$ as the averages of the surface slops as follows:

$$\langle P^+ \rangle = \frac{1}{A_S^+} \int_{S^+} |P^+(x,y)| \, dS^+ = H^+ + W_a^+ + R_a^+$$

$$\langle P^- \rangle = \frac{1}{A_S^-} \int_{S^-} |P^-(x,y)| \, dS^- = H^- + W_a^- + R_a^-$$

$$\langle \partial P^+/\partial \alpha \rangle = \frac{1}{A_S^+} \int_{S^+} \left| \frac{\partial P^+(x,y)}{\partial \alpha} \right| dS^+ = WS_\alpha^+ + RS_\alpha^+ \quad i.e. \ \alpha = x,y \tag{33}$$

$$\langle \partial P^-/\partial \alpha \rangle = \frac{1}{A_S^-} \int_{S^-} \left| \frac{\partial P^-(x,y)}{\partial \alpha} \right| dS^- = WS_\alpha^- + RS_\alpha^- \quad i.e. \ \alpha = x,y$$

where $A_S^+$ and $A_S^-$ denote the surface areas of the upper and lower surfaces of the film, respectively.

In equation (33), $W_a^+$ and $W_a^-$ are introduced as the average waviness of the upper and lower surfaces of the film, respectively. $R_a^+$ and $R_a^-$ denote, respectively, the average roughness of the upper and lower surfaces. These quantities can be defined as follows:

$$R_a^+ = \frac{1}{A_S^+} \int_{S^+} |\mathcal{R}^+(x,y)| \, dS^+$$

$$R_a^- = \frac{1}{A_S^-} \int_{S^-} |\mathcal{R}^-(x,y)| \, dS^-$$

$$W_a^+ = \frac{1}{A_S^+} \int_{S^+} |\varpi^+(x,y)| \, dS^+ \tag{34}$$

$$W_a^- = \frac{1}{A_S^-} \int_{S^+} |\varpi^-(x,y)| \, dS^-$$

In addition to the average waviness and the average roughness, $WS$ and $RS$ are introduced as the averages of the waviness slope and roughness slope, respectively. These averages can be defined as follows:

$$RS_\alpha^+ = \frac{1}{A_S^+} \int_{S^+} \left| \frac{\partial \mathcal{R}^+(x,y)}{\partial \alpha} \right| dS^+ \quad i.e. \ \alpha = x,y$$

$$RS_\alpha^- = \frac{1}{A_S^-} \int_{S^-} \left| \frac{\partial \mathcal{R}^-(x,y)}{\partial \alpha} \right| dS^- \quad i.e. \ \alpha = x,y \tag{35}$$

$$WS_\alpha^+ = \frac{1}{A_S^+} \int_{S+} \left|\frac{\partial \varpi^+(x,y)}{\partial \alpha}\right| dS^+ \quad i.e.\ \alpha = x, y$$

$$WS_\alpha^- = \frac{1}{A_S^-} \int_{S+} \left|\frac{\partial \varpi^-(x,y)}{\partial \alpha}\right| dS^- \quad i.e.\ \alpha = x, y$$

Indeed, the average roughness, $R_a$, and the average waviness, $W_a$, are the common parameters that are usually used to assess the surface textures of a manufactured material. Thus, the experiment usually reports these two parameters. In addition, because of the random profiles of the waviness and the roughness through the surface of a material, the average waviness and average roughness are usually defined with respect to their average slopes, $WS$ and $RS$. Therefore, it is acceptable to depend on these averages to represent the surface waviness and roughness and their slopes in the context of the proposed formulation.

Utilizing the defined averages, $W_a$, $R_a$, $WS$, and $RS$ and neglecting the high-order derivatives of $P(x,y)$, the equation of motion (19) can be written in the following simplified form:

$$\begin{aligned}
D &\left(\frac{\partial^4 w(x,y,t)}{\partial x^4} + 2\frac{\partial^4 w(x,y,t)}{\partial x^2 \partial y^2} + \frac{\partial^4 w(x,y,t)}{\partial y^4}\right) + B_1 \frac{\partial^2 w(x,y,t)}{\partial x^2} + B_2 \frac{\partial^2 w(x,y,t)}{\partial y^2} \\
&+ B_3 \frac{\partial^2 w(x,y,t)}{\partial x \partial y} + K_1 \left(\frac{\partial^2}{\partial x^2} + \frac{\partial^2}{\partial y^2}\right)\frac{\partial w(x,y,t)}{\partial x} \\
&+ K_2 \left(\frac{\partial^2}{\partial x^2} + \frac{\partial^2}{\partial y^2}\right)\frac{\partial w(x,y,t)}{\partial y} - F_z(x,y) + I\ddot{w}(x,y,t) = 0
\end{aligned} \tag{36}$$

where

$$\begin{aligned}
D = &\frac{h^3}{12}(\lambda + 2\mu) + \frac{1}{3}\left(\left(H^+ + W_a^+ + R_a^+ + \frac{h}{2}\right)^3 - \frac{h^3}{8}\right)(\lambda_A^+ + 2\mu_A^+) \\
&+ \frac{1}{3}\left(\left(H^- + W_a^- + R_a^- + \frac{h}{2}\right)^3 - \frac{h^3}{8}\right)(\lambda_A^- + 2\mu_A^-) \\
&+ \left(H^+ + W_a^+ + R_a^+ + \frac{h}{2}\right)^2 (\lambda_S^+ + 2\mu_S^+) \\
&+ \left(H^+ + W_a^+ + R_a^+ + \frac{h}{2}\right)^2 (\lambda_S^- + 2\mu_S^-)
\end{aligned} \tag{37}$$

$$\begin{aligned}
B_1 = &\, 2(\lambda_A^+ + 2\mu_A^+)\left(H^+ + W_a^+ + R_a^+ + \frac{h}{2}\right)(WS_x^+ + RS_x^+)^2 \\
&+ 2(\lambda_A^- + 2\mu_A^-)\left(H^- + W_a^- + R_a^- + \frac{h}{2}\right)(WS_x^- + RS_x^-)^2 \\
&+ 2(\lambda_S^+ + 2\mu_S^+)(WS_x^+ + RS_x^+)^2 + 2(\lambda_S^- + 2\mu_S^-)(WS_x^- + RS_x^-)^2 \\
&+ 2\lambda_A^+ \left(H^+ + W_a^+ + R_a^+ + \frac{h}{2}\right)(WS_y^+ + RS_y^+)^2 \\
&+ 2\lambda_A^- \left(H^- + W_a^- + R_a^- + \frac{h}{2}\right)(WS_y^- + RS_y^-)^2 + 2\lambda_S^+ (WS_y^+ + RS_y^+)^2 \\
&+ 2\lambda_S^- (WS_y^- + RS_y^-)^2
\end{aligned} \tag{38}$$

$$B_2 = 2(\lambda_A^+ + 2\mu_A^+)\left(H^+ + W_a^+ + R_a^+ + \frac{h}{2}\right)(WS_y^+ + RS_y^+)^2$$
$$+ 2(\lambda_A^- + 2\mu_A^-)\left(H^- + W_a^- + R_a^- + \frac{h}{2}\right)(WS_y^- + RS_y^-)^2$$
$$+ 2(\lambda_S^+ + 2\mu_S^+)(WS_y^+ + RS_y^+)^2 + 2(\lambda_S^- + 2\mu_S^-)(WS_y^- + RS_y^-)^2$$
$$+ 2\lambda_A^+\left(H^+ + W_a^+ + R_a^+ + \frac{h}{2}\right)(WS_x^+ + RS_x^+)^2$$
$$+ 2\lambda_A^-\left(H^- + W_a^- + R_a^- + \frac{h}{2}\right)(WS_x^- + RS_x^-)^2 + 2\lambda_S^+(WS_x^+ + RS_x^+)^2$$
$$+ 2\lambda_S^-(WS_x^- + RS_x^-)^2 \tag{39}$$

$$B_3 = 8\mu_A^+\left(H^+ + W_a^+ + R_a^+ + \frac{h}{2}\right)(WS_x^+ + RS_x^+)(WS_y^+ + RS_y^+)$$
$$+ 8\mu_A^-\left(H^- + W_a^- + R_a^- + \frac{h}{2}\right)(WS_x^- + RS_x^-)(WS_y^- + RS_y^-)$$
$$+ 8\mu_S^+(WS_x^+ + RS_x^+)(WS_y^+ + RS_y^+)$$
$$+ 8\mu_S^-(WS_x^- + RS_x^-)(WS_y^- + RS_y^-) \tag{40}$$

$$K_1 = 2(\lambda_A^+ + 2\mu_A^+)\left(H^+ + W_a^+ + R_a^+ + \frac{h}{2}\right)^2 (WS_x^+ + RS_x^+)$$
$$+ 2(\lambda_A^- + 2\mu_A^-)\left(H^- + W_a^- + R_a^- + \frac{h}{2}\right)^2 (WS_x^- + RS_x^-)$$
$$+ 4(\lambda_S^+ + 2\mu_S^+)\left(H^+ + W_a^+ + R_a^+ + \frac{h}{2}\right)(WS_x^+ + RS_x^+)$$
$$+ 4(\lambda_S^- + 2\mu_S^-)\left(H^- + W_a^- + R_a^- + \frac{h}{2}\right)(WS_x^- + RS_x^-) \tag{41}$$

$$K_2 = 2(\lambda_A^+ + 2\mu_A^+)\left(H^+ + W_a^+ + R_a^+ + \frac{h}{2}\right)^2 (WS_y^+ + RS_y^+)$$
$$+ 2(\lambda_A^- + 2\mu_A^-)\left(H^- + W_a^- + R_a^- + \frac{h}{2}\right)^2 (WS_y^- + RS_y^-)$$
$$+ 4(\lambda_S^+ + 2\mu_S^+)\left(H^+ + W_a^+ + R_a^+ + \frac{h}{2}\right)(WS_y^+ + RS_y^+)$$
$$+ 4(\lambda_S^- + 2\mu_S^-)\left(H^- + W_a^- + R_a^- + \frac{h}{2}\right)(WS_y^- + RS_y^-) \tag{42}$$

The boundary conditions which are defined in equation (20) can be rewritten utilizing the introduced average parameters of the surface integrity as presented in Appendix A.

**Case 5:** The equation of motion (33) can be rewritten for an ultra-thin film with the upper and lower surfaces having the same average parameters of the surface integrity, *i.e.* $H^+ = H^- = H$, $W_a^+ = W_a^- = W_a$, $R_a^+ = R_a^- = R_a$, $WS_\alpha^+ = WS_\alpha^- = WS_\alpha$, $RS_\alpha^+ = RS_\alpha^- = RS_\alpha$, $\lambda_A^+ = \lambda_A^- = \lambda_A$, $\mu_A^+ = \mu_A^- = \mu_A$, $\rho_A^+ = \rho_A^- = \rho_A$, $\sigma_s^+ = \sigma_s^- = \sigma_s$, $\lambda_S^+ = \lambda_S^- = \lambda_S$, and $\mu_S^+ = \mu_S^- = \mu_S$, as follows:

$$D\left(\frac{\partial^4 w(x,y,t)}{\partial x^4} + 2\frac{\partial^4 w(x,y,t)}{\partial x^2 \partial y^2} + \frac{\partial^4 w(x,y,t)}{\partial y^4}\right) + B_1 \frac{\partial^2 w(x,y,t)}{\partial x^2} + B_2 \frac{\partial^2 w(x,y,t)}{\partial y^2}$$
$$+ B_3 \frac{\partial^2 w(x,y,t)}{\partial x \partial y} + K_1\left(\frac{\partial^2}{\partial x^2} + \frac{\partial^2}{\partial y^2}\right)\frac{\partial w(x,y,t)}{\partial x} \quad (43)$$
$$+ K_2\left(\frac{\partial^2}{\partial x^2} + \frac{\partial^2}{\partial y^2}\right)\frac{\partial w(x,y,t)}{\partial y} - F_z(x,y) + I\ddot{w}(x,y,t) = 0$$

where

$$D = \frac{h^3}{12}(\lambda + 2\mu) + \frac{2}{3}\left(\left(H + W_a + R_a + \frac{h}{2}\right)^3 - \frac{h^3}{8}\right)(\lambda_A + 2\mu_A)$$
$$+ 2\left(H + W_a + R_a + \frac{h}{2}\right)^2(\lambda_S + 2\mu_S) \quad (44)$$

$$B_1 = 4(\lambda_A + 2\mu_A)\left(H + W_a + R_a + \frac{h}{2}\right)(WS_x + RS_x)^2 + 4(\lambda_S + 2\mu_S)(WS_x + RS_x)^2$$
$$+ 4\lambda_A\left(H + W_a + R_a + \frac{h}{2}\right)(WS_y + RS_y)^2 + 4\lambda_S(WS_y + RS_y)^2 \quad (45)$$

$$B_2 = 4(\lambda_A + 2\mu_A)\left(H + W_a + R_a + \frac{h}{2}\right)(WS_y + RS_y)^2 + 4(\lambda_S + 2\mu_S)(WS_y + RS_y)^2$$
$$+ 4\lambda_A\left(H + W_a + R_a + \frac{h}{2}\right)(WS_x + RS_x)^2 + 4\lambda_S(WS_x + RS_x)^2 \quad (46)$$

$$B_3 = 16\mu_A\left(H + W_a + R_a + \frac{h}{2}\right)(WS_x + RS_x)(WS_y + RS_y)$$
$$+ 16\mu_S(WS_x + RS_x)(WS_y + RS_y) \quad (47)$$

$$K_1 = 4(\lambda_A + 2\mu_A)\left(H + W_a + R_a + \frac{h}{2}\right)^2(WS_x + RS_x)$$
$$+ 8(\lambda_S + 2\mu_S)\left(H + W_a + R_a + \frac{h}{2}\right)(WS_x + RS_x) \quad (48)$$

$$K_2 = 4(\lambda_A + 2\mu_A)\left(H + W_a + R_a + \frac{h}{2}\right)^2(WS_y + RS_y)$$
$$+ 8(\lambda_S + 2\mu_S)\left(H + W_a + R_a + \frac{h}{2}\right)(WS_y + RS_y) \quad (49)$$

$$I = \int_{-h/2}^{h/2} \rho \, dz + 2\int_{h/2}^{h/2 + H + W_a + R_a} \rho_A \, dz \quad (50)$$

The corresponding boundary conditions are explicitly written in Appendix A.

## 4. Selected case study

In order to demonstrate the effects of the surface integrity on the mechanics of ultra-thin films, a selected case study for the cylindrical static bending of a film with infinite width, finite length $L = 200h$, and a thickness $h = 1\mu m$ is considered. The ultra-thin film is made of Al [111] with Young's modulus, $E = 90\ GPa$, and Poisson's ratio, $\nu = 0.3$ [Duan et al., 2005]. The upper and lower surfaces of the ultra-thin film are considered having the same surface parameters (**Case 5**). Each of the altered layers is considered having $\lambda_A = 0.8\lambda$ and $\mu_A = 0.8\mu$ and an average thickness $H = 0 \rightarrow 50\ nm$. To account for the surface topography of the considered film, the following trigonometric function is considered for the surface profile:

$$P(x) = H + A_w \sin\left(\frac{2\pi x}{\lambda_w}\right) + A_r \sin\left(\frac{2\pi x}{\lambda_r}\right) \tag{51}$$

where $A_w$ and $A_r$ denote the amplitudes of the waviness and roughness profiles, respectively. $\lambda_w$ and $\lambda_r$ are, respectively, the waviness and roughness spatial wavelengths along $x-$axis.

Because the considered ultra-thin film possesses identical surfaces, its static equilibrium equation can be obtained from equation (43) as follows:

$$D\frac{d^4w(x)}{dx^4} + B_1 \frac{d^2w(x)}{dx^2} + K_1 \frac{d^3w(x)}{dx^3} - F_z(x) = 0 \tag{52}$$

with

$$D = \frac{h^3}{12}(\lambda + 2\mu) + \frac{2}{3}\left(\left(H + W_a + R_a + \frac{h}{2}\right)^3 - \frac{h^3}{8}\right)(\lambda_A + 2\mu_A) \\ + 2\left(H + W_a + R_a + \frac{h}{2}\right)^2 (\lambda_S + 2\mu_S) \tag{53}$$

$$B_1 = 4(\lambda_A + 2\mu_A)\left(H + W_a + R_a + \frac{h}{2}\right)(WS_x + RS_x)^2 + 4(\lambda_S + 2\mu_S)(WS_x + RS_x)^2 \tag{54}$$

$$K_1 = 4(\lambda_A + 2\mu_A)\left(H + W_a + R_a + \frac{h}{2}\right)^2 (WS_x + RS_x) \\ + 8(\lambda_S + 2\mu_S)\left(H + W_a + R_a + \frac{h}{2}\right)(WS_x + RS_x) \tag{55}$$

It should be noted that, according to the defined surface profile function in equation (51), the average parameters of the surface integrity presented in equations (52)-(55) can be obtained as follows:

$$R_a = \frac{A_r}{L}\int_0^L \left|\sin\left(\frac{2\pi x}{\lambda_r}\right)\right| dx \\ W_a = \frac{A_w}{L}\int_0^L \left|\sin\left(\frac{2\pi x}{\lambda_w}\right)\right| dx \tag{56}$$

$$RS_x = \frac{2\pi A_r}{\lambda_r L} \int_0^L \left|\cos\left(\frac{2\pi x}{\lambda_r}\right)\right| dx$$

$$WS_x = \frac{2\pi A_w}{\lambda_w L} \int_0^L \left|\cos\left(\frac{2\pi x}{\lambda_w}\right)\right| dx$$

Utilizing $X = \frac{x}{L}$ and $W(X) = \frac{w(x)}{h}$ as nondimensional parameters, equation (52) can be written in a normalized form as follows:

$$\frac{d^4 W(X)}{dX^4} + S_1 \frac{d^2 W(X)}{dX^2} + S_2 \frac{d^3 W(X)}{dX^3} = q_0 \tag{57}$$

where

$$S_1 = \frac{B_1 L^2}{D}$$

$$S_2 = \frac{K_1 L}{D} \tag{58}$$

$$q_0 = \frac{F_z L^4}{Dh}$$

where $F_z$ is the transverse applied force per unit surface area. In this study, the ultra-thin film is subjected to a non-uniform distributed load such that $F_z = KX^2$, i.e. $K$ is the forcing amplitude.

### 4.1. Analytical solution

For the considered loading case, a general solution for equation (57) can be derived as follows:

$$W(X) = C_1 + C_2 X + e^{-\frac{S_2}{2}X}\left(C_3 \cos\left(\left(\frac{1}{2}\sqrt{4S_1 - S_2^2}\right)X\right) + C_4 \sin\left(\left(\frac{1}{2}\sqrt{4S_1 - S_2^2}\right)X\right)\right) \\ + \left(\frac{KL^4}{12DhS_1}\right)X^4 - \left(\frac{KL^4 S_2}{3DhS_1^2}\right)X^3 + \left(\frac{KL^4 S_2^2}{DhS_1^3} - \frac{KL^4}{DhS_1^2}\right)X^2 \tag{59}$$

According to the defined boundary conditions in equation (A.8), for a clamped-clamped film, the following boundary conditions are applied:

$$W(0) = 0, \frac{dW(0)}{dX} = 0$$
$$W(1) = 0, \frac{dW(1)}{dX} = 0 \tag{60}$$

Utilizing these boundary conditions, the constants $C_i$ are obtained as presented in Appendix B.

When the surface integrity effects are eliminated from the proposed formulation, equation (57) reduces to (**Case 4**):

$$\frac{d^4 W(X)}{dX^4} = q_0 \tag{61}$$

A general solution for equation (61) can be obtained for the considered non-uniform distributed load as follows:

$$W(X) = \left(\frac{KL^4}{Dh}\right)\left(\frac{X^6}{360} - \frac{X^3}{90} + \frac{X^2}{120}\right) \tag{62}$$

### 4.2. Parametric study

Effects of the surface roughness, surface waviness, altered layers, and film size on the static bending of ultra-thin films are investigated. To this end, the nondimensional deflection distribution along the film length is depicted for different values of the film thickness, the roughness and waviness wavelengths and amplitudes, and the altered layers average thickness. In fact, effects of the surface excess energy have been intensively investigated in various studies [He et al., 2004; Sharma and Ganti, 2004; He and Li, 2006; Zhou and Huang, 2004; Shim et al., 2005; Sun and Zhang, 2003; Zhang and Sun, 2004; Guo and Zhao, 2005; Lu et al., 2006; Huang, 2008; Shaat et al., 2013a, 2013b, 2014; Shaat and Abdelkefi, 2017b]. Therefore, in the present study, we focus on investigating the effects of the other parameters of the surface integrity where $\lambda_S = 0$ and $\mu_S = 0$ are considered.

*Surface roughness effects*

Effects of the wavelength and amplitude of the surface roughness are depicted in Figures 2-4, respectively. In these figures, the results of the proposed formulation (equation (57)) are compared to the results of the classical formulation of thin-films (equation (61)) with No Surface Integrity (NSI) effects. It is clear that the ultra-thin film is sensitive to the small changes in the surface roughness. The ultra-thin film reflects different deflections for the different values of the roughness wavelength and amplitude.

Figure 2 shows the effects of the wavelength of the surface roughness on the deflection of the ultra-thin film when the waviness and altered layers effects are neglected, *i.e.* $A_w = 0$ and $H = 0$. A small wavelength indicates a rough surface with sharp asperities while a large wavelength refers to a smooth surface with a surface waviness. Different nondimensional deflections are obtained for the different values of the roughness wavelength. For example, the nondimensional maximum deflection corresponding to a wavelength $\lambda_r = 0.05L$ is $\sim 2.25$ times the one corresponding to a wavelength $\lambda_r = 0.01L$. Furthermore, as the roughness wavelength is increased, the nondimensional deflection distribution merges with the obtained nondimensional deflection distribution when all the surface integrity effects are neglected. Under the considered loading conditions, an ultra-thin film with strong roughness (small $\lambda_r$) deforms such that its maximum deflection shifts to the left and decreases with the decrease in the roughness wavelength value.

The effects of the roughness wavelength when the surface roughness effects are coupled with the waviness and altered layers effects are depicted in Figure 3. When the measures for the effects of the surface waviness and the altered layers are considered, different deflections than the ones plotted in Figure 2 are obtained. For instance, the incorporation of the surface waviness and altered layers effects increases the nondimensional maximum deflection of a thin film with $\lambda_r = 0.01L$ to $\sim 1.65$ times its depicted value in Figure 2. Because of the incorporated waviness and altered layers effects, merging with the NSI deflection curve is not an option for all the roughness wavelength values.

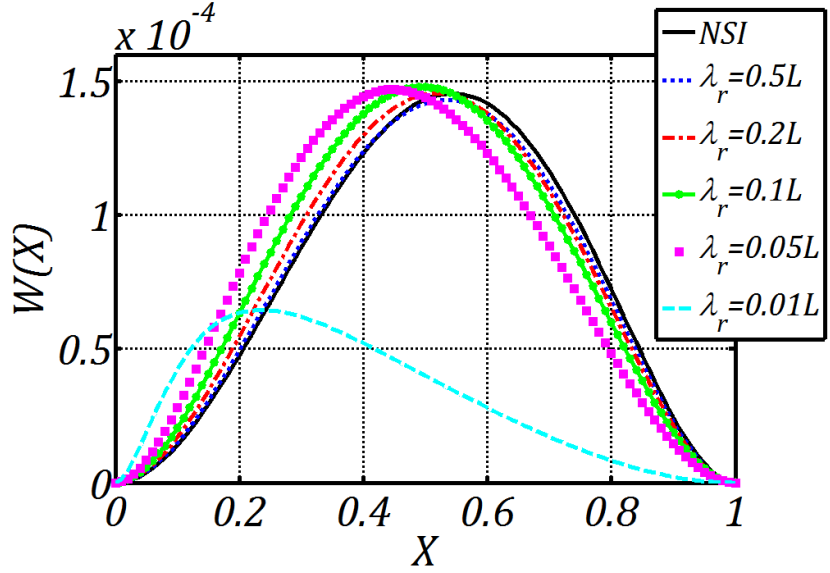

Figure 2: Effects of the wavelength of the surface roughness on the static bending of the ultra-thin film under a non-uniform distributed load of an amplitude $K = 1$. ($A_r = 10\ nm$, $A_w = 0$, and $H = 0$). *NSI: No Surface Integrity effects are considered.*

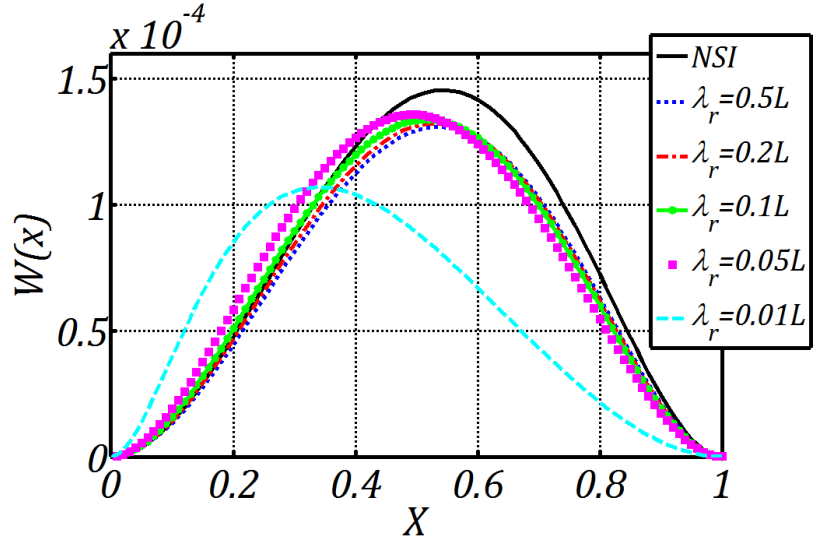

Figure 3: Effects of the wavelength of the surface roughness on the static bending of the ultra-thin film under a non-uniform distributed load of an amplitude $K = 1$. ($\lambda_w = 0.5L$, $A_w = 2\ nm$, $A_r = 5\ nm$, and $H = 20\ nm$).

Figure 4 shows the effects of the roughness amplitude on the static bending deflection of the ultra-thin film. Different nondimensional deflections are obtained for the different amplitude values. Films with strong rough surfaces (with large $A_r$) reflect smaller deflections than films with smooth surfaces (*i.e.* $A_r = 0$). The deformed configuration of the film bends to the left due to the surface roughness effects. Thus, the nondimensional maximum deflection shifts to the left with the increase in the amplitude value.

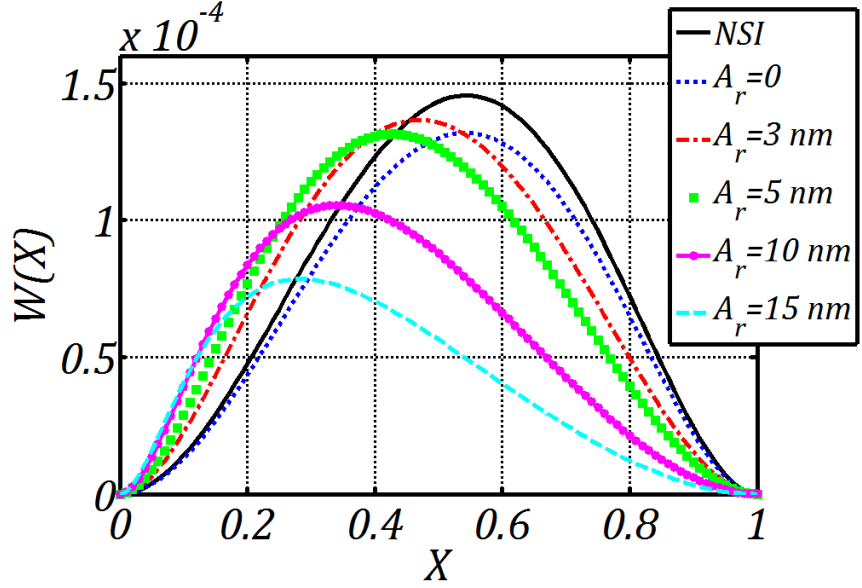

Figure 4: Effects of the amplitude of the surface roughness on the static bending of the ultra-thin film under a non-uniform distributed load of an amplitude $K = 1$. ($\lambda_w = 0.5L$, $A_w = 2\ nm$, $\lambda_r = 0.02L$, and $H = 20\ nm$).

According to the plotted curves in Figures 2-4, it can be demonstrate that the real profile of the surface roughness should be considered to guarantee accurate predictions of the mechanics of ultra-thin films. it is revealed that the mechanics of ultra-thin films is highly sensitive to their surface roughness. Moreover, these figures reflect the hardening effects of the surface roughness. It is clear that films with rough surfaces are more rigid than those of smooth surfaces.

*Surface waviness effects*

As for the effects of the surface waviness, the film deflection is presented for the different waviness wavelength and amplitude values in Figures 5 and 6. The plotted curves in these two figures are obtained when the surface roughness and altered layers effects are incorporated. According to the plotted curves in Figure 5, the waviness wavelength has a negligible effect on the static deflection of ultra-thin films. As shown in the figure, the thin-film reflects the same nondimensional deflection distribution for all the wavelength values which is coincident with the nondimensional deflection distribution when the waviness effects are eliminated. On the other hand, Figure 6 reflects the effects of the waviness amplitude on the film deflection. The increase in the waviness amplitude is accompanied with a decrease in the film 's maximum deflection. This indicates that the surface waviness affect the mechanics of ultra-thin films with a hardening mechanism.

*Altered layers effects*

The plotted curves in Figure 7 show the effects of the altered layers on the static deflection of the ultra-thin film. In this figure, the nondimensional deflection distribution is plotted for the different values of the altered layers' average thickness. Inspecting the plotted curves in Figure 7, it clear that the altered layers, as well, affect the film with a hardening mechanism where the increase in their average thickness is accompanied with a decrease in the film's maximum deflection.

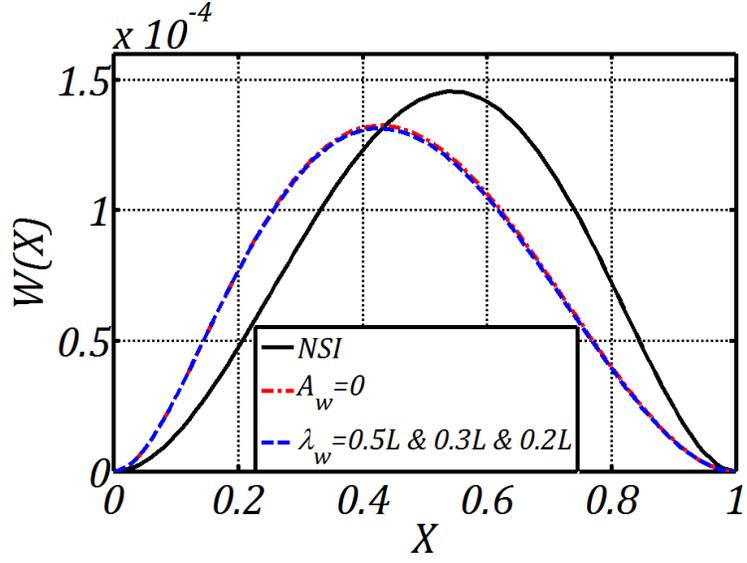

Figure 5: Effects of the wavelength of the surface waviness on the static bending of the ultra-thin film under a non-uniform distributed load of an amplitude $K = 1$. ($A_w = 2\ nm$, $\lambda_r = 0.02L$, $A_r = 5\ nm$, and $H = 20\ nm$).

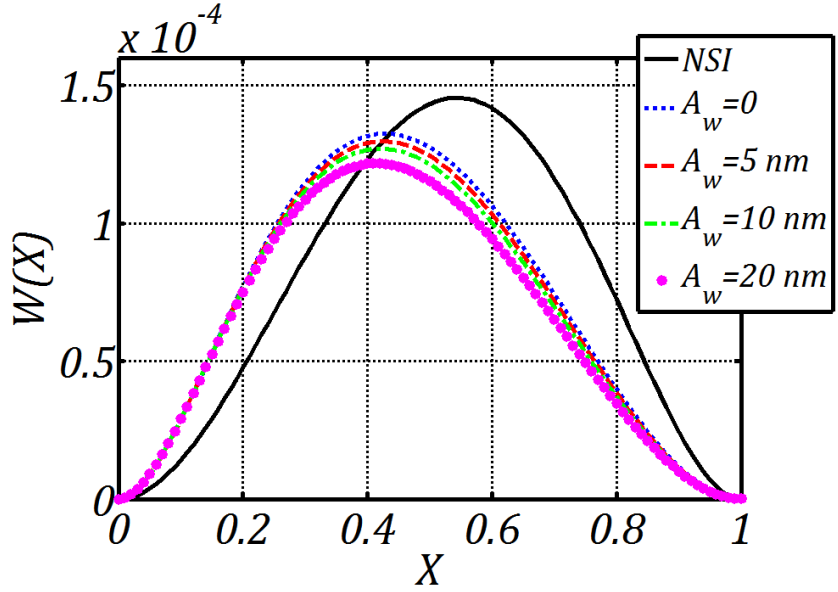

Figure 6: Effects of the amplitude of the surface waviness on the static bending of the ultra-thin film under a non-uniform distributed load of an amplitude $K = 1$. ($\lambda_w = 0.5L$, $\lambda_r = 0.02L$, $A_r = 5\ nm$, and $H = 20\ nm$).

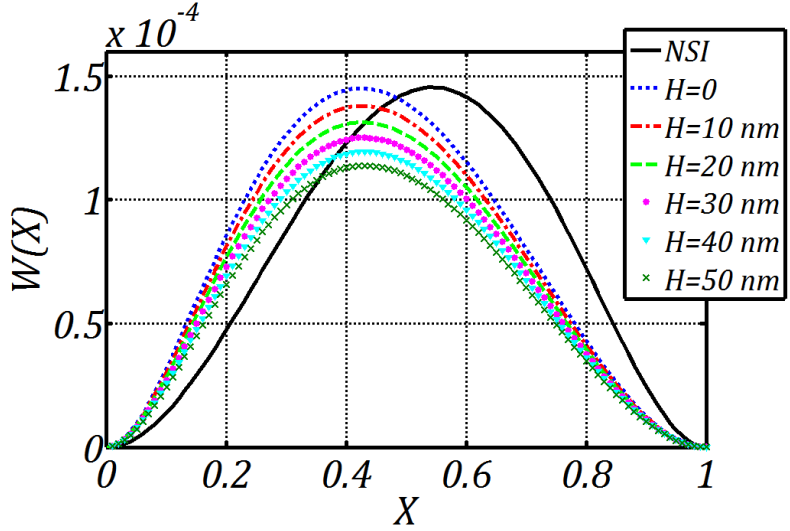

Figure 7: Effects of the altered layers on the static bending of the ultra-thin film under a non-uniform distributed load of an amplitude $K = 1$. ($\lambda_w = 0.5L$, $A_w = 2\,nm$, $\lambda_r = 0.02L$, and $A_r = 5\,nm$).

*Film size effects*

The curves in Figure 8 are plotted to reveal the size-dependent behavior of the ultra-thin film. As shown in this figure, different nondimensional deflections are obtained for the different film thickness values. The decrease in the film thickness is accompanied with a decrease in the film deflection. This can be attributed to the increase in the surface integrity effects, which affect the mechanics of the ultra-thin film with a hardening mechanism, with the decrease in the film thickness. It can be concluded that, the mechanics of ultra-thin films are significantly altered because of the surface integrity. However, the surface integrity has a negligible effects on the elastic behaviors of conventional films with large sizes.

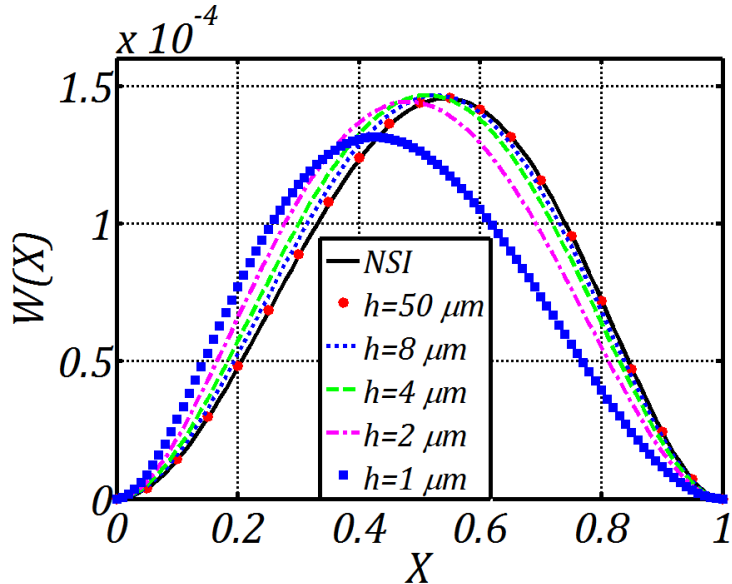

Figure 8: Effects of the ultra-thin film thickness on its static bending. ($\lambda_w = 0.5L$, $A_w = 2\,nm$, $\lambda_r = 0.02L$, $A_r = 5\,nm$, $H = 20\,nm$, $K = 1$, and $L = 200h$).

## 5. Conclusions

A detailed formulation for the effects of the surface integrity on the mechanics of ultra-thin films was presented in the framework of linear elasticity. In this formulation, the ultra-thin film is modeled as a material bulk covered with two altered layers and two outermost surfaces as distinct phases. Two different surface profile functions were introduced for the upper and lower surfaces to model their surface topography. Two versions of the proposed formulation were presented where the governing equations were derived depending on the general form of the surface profile functions in one version and depending on the average of the surface profiles in another version. These two formulations incorporate the essential measures which are needed to account for the effects of the surface topography, *i.e.* waviness and roughness, surface metallurgy, *i.e.* altered layer, and surface excess energy, *i.e.* surface stress and energy, on the mechanics of ultra-thin films.

A selected case study for the static bending of a clamped-clamped ultra-thin film was analytically solved to investigate the effects the surface integrity on the mechanics of ultra-thin films. An intensive study on the effects of the surface roughness, surface waviness, altered layers, and film size on the static bending of ultra-thin films was presented. In this study new size-dependent behaviors of ultra-thin films which depend on the surface integrity were revealed. The obtained results reflected the significant impacts of the surface integrity on the mechanics of ultra-thin films. Thus, the mechanics of ultra-thin films can be significantly altered for any small variations in the surface roughness, surface waviness, or altered layers properties.

## Appendix A

Utilizing the average parameters of the surface integrity, $W_a$, $R_a$, $WS$, and $RS$, introduced in equations (34) and (35), the boundary conditions (equation (20)) can be rewritten as follows:

On edge $\Gamma_x$:

$$-(D_1 + D_2)(w_{,xxx} + w_{,yyx}) - \frac{1}{2}(A_1 + A_2)w_{,xx} - A_2 w_{,yy} - A_3 w_{,xy} + \sigma_S^+(WS_x^+ + RS_x^+) - \sigma_S^-(WS_x^- + RS_x^-) = \bar{V}_x \text{ or } w = w^{\Gamma_x}$$

$$-(D_1 + D_2)w_{,xx} - D_2 w_{,yy} + \sigma_S^+(h/2 + H^+ + W_a^+ + R_a^+) - \sigma_S^-(h/2 + H^- + W_a^- + R_a^-) = \bar{M}_{xx} \text{ or } w_{,x} = w_{,x}^{\Gamma_x}$$

(A.1)

On edge $\Gamma_y$:

$$-(D_1 + D_2)(w_{,yyy} + w_{,xxy}) - \frac{1}{2}(A_3 + A_4)w_{,yy} - A_4 w_{,xx} - A_1 w_{,xy} + \sigma_S^+(WS_y^+ + RS_y^+) - \sigma_S^-(WS_y^- + RS_y^-) = \bar{V}_y \text{ or } w = w^{\Gamma_y}$$

$$-(D_1 + D_2)w_{,yy} - D_2 w_{,xx} + \sigma_S^+\left(\frac{h}{2} + H^+ + W_a^+ + R_a^+\right) - \sigma_S^-\left(\frac{h}{2} + H^- + W_a^- + R_a^-\right) = \bar{M}_{yy} \text{ or } w_{,y} = w_{,y}^{\Gamma_y}$$

where

$$D_1 = \mu\left(\frac{h^3}{6}\right) + \frac{2\mu_A^+}{3}\left(\left(H^+ + W_a^+ + R_a^+ + \frac{h}{2}\right)^3 - \frac{h^3}{8}\right)$$

$$+ \frac{2\mu_A^-}{3}\left(\left(H^- + W_a^- + R_a^- + \frac{h}{2}\right)^3 - \frac{h^3}{8}\right) \quad \text{(A.2)}$$

$$+ 2\mu_S^+\left(H^+ + W_a^+ + R_a^+ + \frac{h}{2}\right)^2 + 2\mu_S^-\left(H^+ + W_a^+ + R_a^+ + \frac{h}{2}\right)^2$$

$$D_2 = \lambda\left(\frac{h^3}{12}\right) + \frac{\lambda_A^+}{3}\left(\left(H^+ + W_a^+ + R_a^+ + \frac{h}{2}\right)^3 - \frac{h^3}{8}\right)$$

$$+ \frac{\lambda_A^-}{3}\left(\left(H^- + W_a^- + R_a^- + \frac{h}{2}\right)^3 - \frac{h^3}{8}\right) + \lambda_S^+\left(H^+ + W_a^+ + R_a^+ + \frac{h}{2}\right)^2 \quad \text{(A.3)}$$

$$+ \lambda_S^-\left(H^+ + W_a^+ + R_a^+ + \frac{h}{2}\right)^2$$

$$A_1 = 2\mu_A^+\left(H^+ + W_a^+ + R_a^+ + \frac{h}{2}\right)^2 (WS_x^+ + RS_x^+)$$

$$+ 2\mu_A^-\left(H^- + W_a^- + R_a^- + \frac{h}{2}\right)^2 (WS_x^- + RS_x^-)$$

$$+ 4\mu_S^+\left(H^+ + W_a^+ + R_a^+ + \frac{h}{2}\right)(WS_x^+ + RS_x^+) \quad \text{(A.4)}$$

$$+ 4\mu_S^-\left(H^- + W_a^- + R_a^- + \frac{h}{2}\right)(WS_x^- + RS_x^-)$$

$$A_2 = \lambda_A^+\left(H^+ + W_a^+ + R_a^+ + \frac{h}{2}\right)^2 (WS_x^+ + RS_x^+)$$

$$+ \lambda_A^-\left(H^- + W_a^- + R_a^- + \frac{h}{2}\right)^2 (WS_x^- + RS_x^-)$$

$$+ 2\lambda_S^+\left(H^+ + W_a^+ + R_a^+ + \frac{h}{2}\right)(WS_x^+ + RS_x^+) \quad \text{(A.5)}$$

$$+ 2\lambda_S^-\left(H^- + W_a^- + R_a^- + \frac{h}{2}\right)(WS_x^- + RS_x^-)$$

$$A_3 = 2\mu_A^+\left(H^+ + W_a^+ + R_a^+ + \frac{h}{2}\right)^2 (WS_y^+ + RS_y^+)$$

$$+ 2\mu_A^-\left(H^- + W_a^- + R_a^- + \frac{h}{2}\right)^2 (WS_y^- + RS_y^-)$$

$$+ 4\mu_S^+\left(H^+ + W_a^+ + R_a^+ + \frac{h}{2}\right)(WS_y^+ + RS_y^+) \quad \text{(A.6)}$$

$$+ 4\mu_S^-\left(H^- + W_a^- + R_a^- + \frac{h}{2}\right)(WS_y^- + RS_y^-)$$

$$A_4 = \lambda_A^+ \left(H^+ + W_a^+ + R_a^+ + \frac{h}{2}\right)^2 (WS_y^+ + RS_y^+)$$

$$+ \lambda_A^- \left(H^- + W_a^- + R_a^- + \frac{h}{2}\right)^2 (WS_y^- + RS_y^-)$$

$$+ 2\lambda_S^+ \left(H^+ + W_a^+ + R_a^+ + \frac{h}{2}\right)(WS_y^+ + RS_y^+)$$

$$+ 2\lambda_S^- \left(H^- + W_a^- + R_a^- + \frac{h}{2}\right)(WS_y^- + RS_y^-)$$
(A.7)

**Case 5:** For an ultra-thin film with the upper and lower surfaces having the same average parameters of the surface integrity, *i.e.* $H^+ = H^- = H$, $W_a^+ = W_a^- = W_a$, $R_a^+ = R_a^- = R_a$, $WS_\alpha^+ = WS_\alpha^- = WS_\alpha$, $RS_\alpha^+ = RS_\alpha^- = RS_\alpha$, $\lambda_A^+ = \lambda_A^- = \lambda_A$, $\mu_A^+ = \mu_A^- = \mu_A$, $\rho_A^+ = \rho_A^- = \rho_A$, $\sigma_s^+ = \sigma_s^- = \sigma_s$, $\lambda_S^+ = \lambda_S^- = \lambda_S$, and $\mu_S^+ = \mu_S^- = \mu_S$, the boundary conditions (equation (A.1)) can be rewritten as follows:

On edge $\Gamma_x$:
$$-(D_1 + D_2)(w_{,xxx} + w_{,yyx}) - \frac{1}{2}(A_1 + A_2)w_{,xx} - A_2 w_{,yy} - A_3 w_{,xy} = \bar{V}_x \text{ or } w = w^{\Gamma_x}$$
$$-(D_1 + D_2)w_{,xx} - D_2 w_{,yy} = \bar{M}_{xx} \text{ or } w_{,x} = w_{,x}^{\Gamma_x}$$

On edge $\Gamma_y$:
$$-(D_1 + D_2)(w_{,yyy} + w_{,xxy}) - \frac{1}{2}(A_3 + A_4)w_{,yy} - A_4 w_{,xx} - A_1 w_{,xy} = \bar{V}_y \text{ or } w = w^{\Gamma_y}$$
$$-(D_1 + D_2)w_{,yy} - D_2 w_{,xx} = \bar{M}_{yy} \text{ or } w_{,y} = w_{,y}^{\Gamma_y}$$
(A.8)

where

$$D_1 = \mu\left(\frac{h^3}{6}\right) + \frac{4\mu_A}{3}\left(\left(H + W_a + R_a + \frac{h}{2}\right)^3 - \frac{h^3}{8}\right) + 4\mu_S\left(H + W_a + R_a + \frac{h}{2}\right)^2$$
(A.9)

$$D_2 = \lambda\left(\frac{h^3}{12}\right) + \frac{2\lambda_A}{3}\left(\left(H + W_a + R_a + \frac{h}{2}\right)^3 - \frac{h^3}{8}\right) + \lambda_S\left(H + W_a + R_a + \frac{h}{2}\right)^2$$
(A.10)

$$A_1 = 4\mu_A\left(H + W_a + R_a + \frac{h}{2}\right)^2 (WS_x + RS_x)$$
$$+ 8\mu_S\left(H + W_a + R_a + \frac{h}{2}\right)(WS_x + RS_x)$$
(A.11)

$$A_2 = 2\lambda_A\left(H + W_a + R_a + \frac{h}{2}\right)^2 (WS_x + RS_x)$$
$$+ 4\lambda_S\left(H + W_a + R_a + \frac{h}{2}\right)(WS_x + RS_x)$$
(A.12)

$$A_3 = 4\mu_A\left(H + W_a + R_a + \frac{h}{2}\right)^2 (WS_y + RS_y)$$
$$+ 8\mu_S\left(H + W_a + R_a + \frac{h}{2}\right)(WS_y + RS_y)$$
(A.13)

$$A_4 = 2\lambda_A\left(H + W_a + R_a + \frac{h}{2}\right)^2 (WS_y + RS_y)$$
$$+ 4\lambda_S\left(H + W_a + R_a + \frac{h}{2}\right)(WS_y + RS_y)$$
(A.14)

## Appendix B

By applying the boundary conditions in equation (60), the constants $C_i$ of the film deflection (equation (59)) can be obtained as follows:

$$C_4 = \frac{Q_1\eta_2 - Q_2\eta_4}{\eta_1\eta_2 - \eta_3\eta_4}$$

$$C_3 = \frac{-Q_1 + C_4\eta_1}{\eta_4} \tag{A.15}$$

$$C_2 = \frac{S_2}{2}C_3 - \frac{C_4}{2}\sqrt{4S_1 - S_2^2}$$

$$C_1 = -C_3$$

with

$$\eta_1 = \frac{1}{2}\sqrt{4S_1 - S_2^2} - e^{-\frac{S_2}{2}}\sin\left(\frac{1}{2}\sqrt{4S_1 - S_2^2}\right)$$

$$\eta_2 = \frac{S_2}{2} - e^{-\frac{S_2}{2}}\left(\frac{1}{2}\sqrt{4S_1 - S_2^2}\right)\sin\left(\frac{1}{2}\sqrt{4S_1 - S_2^2}\right) - \frac{S_2}{2}e^{-\frac{S_2}{2}}\cos\left(\frac{1}{2}\sqrt{4S_1 - S_2^2}\right)$$

$$\eta_3 = \frac{1}{2}\sqrt{4S_1 - S_2^2} - e^{-\frac{S_2}{2}}\left(\frac{1}{2}\sqrt{4S_1 - S_2^2}\right)\cos\left(\frac{1}{2}\sqrt{4S_1 - S_2^2}\right) \tag{A.16}$$

$$+ \frac{S_2}{2}e^{-\frac{S_2}{2}}\sin\left(\frac{1}{2}\sqrt{4S_1 - S_2^2}\right)$$

$$\eta_4 = 1 - \frac{S_2}{2} + e^{-\frac{S_2}{2}}\cos\left(\frac{1}{2}\sqrt{4S_1 - S_2^2}\right)$$

## References

Astakhov VP. Surface integrity-definition and importance in functional performance. in Davim JP, (Ed.) Surface Integrity in Machining. SpringerVerlag, London, 2010.

Bellows G, Tishler N, Introduction to surface integrity. GEC Rep., 1970.

Cammarata RC. Surface and interface stress effects in thin-films. Progress in Surface Science 1994; 46(1):1–38.

Cammarata RC, Sieradzki K. Effects of surface stress on the elastic moduli of thin films and superlattices. Physical Review Letters 1989; 62:2005–2008.

Craighead HG. Nanoelectromechanical systems. Science 2000; 290:1532–1535.

Duan HL, Wang J, Huang ZP, Karihaloo BL. Size-dependent effective elastic constants of solids containing nano-inhomogeneities with interface stress. Journal of the Mechanics and Physics of Solids 2005;53:1574–1596.

Duan H, Xue Y, Yi X. Vibration of cantilevers with rough surfaces. Acta Mechanica Solida Sinica, 2009; 22(6):550-554.

Dieter GE. Mechanical Metallurgy. McGraw-Hill, New York, 1988.

Fang X-Q, Zhu C-S. Size-dependent nonlinear vibration of nonhomogeneous shell embedded with a piezoelectric layer based on surface/interface theory. Composite Structures 2017; 160:1191–1197.

Fishman G, Calecki D. Influence of surface roughness on the conductivity of metallic and semiconducting quasi-2-dimensional structures. Physics Review B 1991; 43(14):11851-11855.

Greenwood JA, Williamson JBP. Contact of nominally flat surfaces. Proceedings of the Royal Society of London. Series A-Mathematical and Physical Sciences 1966; 295(1442):300-319.

Guo JG, Zhao YP. The size-dependent elastic properties of nanofilms with surface effects. Journal of Applied Physics 2005; 98 (7):074306.

He LH, Lim CW, Wu BS. A continuum model for size-dependent deformation of elastic films of nano-scale thickness. International Journal of Solids and Structures 2004; 41(3–4):847–857.

He LH, Li ZR. Impact of surface stress on stress concentration. International Journal of Solids and Structures 2006; 43(20):6208–6219.

Huang DW. Size-dependent response of ultra-thin films with surface effects. International Journal of Solids and Structures 2005; 45:568–579.

Huang Q, Ren JX. Surface integrity and its effects on the fatigue life of the nickel-based superalloy GH33A. Int J Fatigue 1991; 13(4):322-326.

Ilic B, Yang Y, Craighead HG. Virus detection using nanoelectromechanical devices. Applied Physics Letters 2004; 85:2604-2606.

Li M, Wang G-C, Min H-G. Effect of surface roughness on magnetic properties of Co films on plasma-etched Si(100) substrates. Journal of Applied Physics 1998; 83(10): 5313-5320.

Llanes L, Casas B, Idanez E, Marsal M, Anglada M. Surface integrity effects on the fracture resistance of electrical-discharge-machined WC–Co cemented carbides. J. Am. Ceram. Soc., 2004; 87(9):1687–1693.

Lu P, He LH, Lu C. Thin plate theory including surface effects. Int. J. Solids Struct. 2006; 43(16):4631-4647.

Lü CF, Lim CW, Chen WQ. Size-dependent elastic behavior of FGM ultra-thin films based on generalized refined theory. Int. J. Solids Struct. 2009;46:1176-1185.

Mohr M, Caron A, Engel PH, Bennewitz R, Gluche P, Brühne K, Fecht HJ. Young's modulus, fracture strength, and Poisson's ratio of nanocrystalline diamond films. Journal of Applied Physics 2014; 116:124308.

Muller P, Saul A. Elastic effects on surface physics. Surface Science Reports 2004; 54(5–8): 157–258.

Peressadko AG, Hosoda N, Persson BNJ. Influence of surface roughness on adhesion between elastic bodies. Physical Review Letters 2005; 95(12):124301-1-4.

Raghu P, Preethi K, Rajagopal A, Reddy JN. Nonlocal third-order shear deformation theory for analysis of laminated plates considering surface stress effects. Composite Structures, 2016; 139:13–29.

Ramulu M, Paul G, Patel J. EDM surface effects on the fatigue strength of a 15 vol% SiCp/Al metal matrix composite material. Composite Structures 2001; 54(1):79–86

Silva ECCM, Tong L, Yip S, Vliet KJV. Size effects on the stiffness of silica nanowires. Small 2006; 2(2):239.

Sinnott MM, Hoeppner DW, Romney E, Dew PA. Effects of surface integrity on the fatigue life of thin flexing membranes. ASAIO Trans. 1989; 35(3):687-90.

Shaat M, Mahmoud FF, Alieldin SS, Alshorbagy AE. Bending analysis of ultra-thin functionally graded Mindlin plates incorporating surface energy effects. International Journal of Mechanical Sciences 2013a; 75:223–232.

Shaat M, Mahmoud FF, Alieldin SS, Alshorbagy AE. Finite element analysis of functionally graded nano-scale films. Finite Element in Analysis and Design 2013b; 74:41–52.

Shaat M, Mahmoud FF, Gao X-L, Faheem AF. Size-dependent bending analysis of Kirchhoff nano-plates based on a modified couple-stress theory including surface effects" International Journal of Mechanical Sciences 2014; 79:31-37.

Shaat M. Effects of grain size and microstructure rigid rotations on the bending behavior of nanocrystalline material beams. International Journal of Mechanical Sciences 2015; 94-95:27–35.

Shaat M, Abdelkefi A. Modeling the material structure and couple stress effects of nanocrystalline silicon beams for pull-in and bio-mass sensing applications. International Journal of Mechanical Sciences 2015; 101-102:280-291.

Shaat M, Abdelkefi A. Reporting buckling strength and elastic properties of nanowires. J. Appl. Phys. 2016; 120:235104.

Shaat M, Abdelkefi A. Buckling characteristics of nanocrystalline nanobeams. Int J Mech Mater Des 2017a; DOI 10.1007/s10999-016-9361-2.

Shaat M Abdelkefi A, Material structure and size effects on the nonlinear dynamics of electrostatically-actuated nano-beams. International Journal of Non-Linear Mechanics 2017b; 89: 25–42.

Sharma P, Ganti S. Size-dependent Eshelby's tensor for embedded nano-inclusions incorporating surface/interface energies. Journal of Applied Mechanics-Transactions of the ASME 2004; 71(5):663–671.

Sharman ARC, Aspinwall DK, Dewes RC, Clifton D, Bowen P. The effects of machined workpiece surface integrity on the fatigue life of γ-titanium aluminide. International Journal of Machine Tools & Manufacture 2001; 41:1681–1685.

Shim HW, Zhou LG, Huang HC, Cale TS. Nanoplate elasticity under surface reconstruction. Applied Physics Letters 2005; 86(15):151912.

Sun CT, Zhang HT. Size-dependent elastic moduli of platelike nanomaterials. Journal of Applied Physics 2003; 93(2):1212–1218.

Wang W, Li P, Jin F, Wang J. Vibration analysis of piezoelectric ceramic circular nanoplates considering surface and nonlocal effects. Composite Structures 2016;140:758–775

Weissmuller J, Duan H. Cantilever Bending with Rough Surfaces. Physical Review Letters 2008; 101:146102.

Yang C, Tartaglino U, Persson BN. Influence of surface roughness on superhydrophobicity. Physics Review Letters 2006; 97(11):116103-1-4.

Zaltin N, Field M. Procedures and precautions in machining titanium alloys. Titanium Sci. Technol. 1973; 1:489–504.

Zhang HT, Sun CT. Nanoplate model for platelike nanomaterials. AIAA Journal 2004; 42(10): 2002–2009.

Zhang H, Lamb RN, Cookson DJ. Nanowetting of rough superhydrophobic surfaces. Applied Physics Letters 2007; 91(25): 254106-1-3.

Zhou LG, Huang HC. Are surfaces elastically softer or stiffer? Applied Physics Letters 2004; 84: 1940–1942.